\documentclass{article}
\usepackage{amsmath}
\usepackage{xcolor}
\usepackage{authblk}

\usepackage{siunitx}
\usepackage{amsmath}

\usepackage[sort,square,numbers]{natbib}

\usepackage{url}

\usepackage{graphicx}

\graphicspath{{Images/}}

\begin{document}
	
	\title{\textbf{Dispersive coupling between $\text{MoSe}_{2}$ and a zero-dimensional integrated nanocavity}}
	\author[1]{David Rosser}
	\author[2]{Dario Gerace}
	\author[3]{Yueyang Chen}
	\author[1]{Yifan Liu}
	\author[3]{James Whitehead}
	\author[3]{Albert Ryou}
	\author[2]{Lucio C. Andreani}
	\author[1,3,*]{Arka Majumdar}
	\affil[1]{Department of Physics, University of Washington, Seattle, Washington 98195, USA}
	\affil[2]{Dipartimento di Fisica, Universit\`{a} di Pavia, 27100 Pavia}
	\affil[3]{Department of Electrical and Computer Engineering, University of Washington, Seattle, Washington 98195, USA}
	\affil[*]{Corresponding Author: arka@uw.edu}
	
	\maketitle
	
	\begin{abstract}
		Establishing a coherent interaction between a material resonance and an optical cavity is a necessary first step for the development of semiconductor quantum optics. Here we demonstrate a coherent interaction between the neutral exciton in monolayer $\text{MoSe}_{2}$ and a zero-dimensional, small mode volume nanocavity. This is observed through a dispersive shift of the cavity resonance when the exciton-cavity detuning is decreased, with an estimated exciton-cavity coupling of $\sim 4.3$ \si{\milli\electronvolt} and a cooperativity of $C \sim 3.4$ at 80 \si{\kelvin}. This coupled exciton-cavity platform is expected to reach the strong light-matter coupling regime (i.e., with $C \sim 380$) at 4 \si{\kelvin} for applications in quantum or ultra-low power nanophotonics.
	\end{abstract}
	
	\section*{Introduction}
	Atomically thin van der Waals (vdW) materials coupled with nanophotonic structures have recently emerged as a promising platform for hybrid integrated photonics \cite{liu_van_2019} due to their easy integration onto a substrate via vdW forces without concern for lattice matching. In particular, the large excitonic binding energy of transition metal dichalcogenides (TMDs) presents an excellent opportunity to study the quantum light-matter interaction in large-scale photonic systems. The first step towards such a quantum system is to demonstrate a coherent interaction between a small mode volume cavity and TMD excitons. To that end, several research groups have observed exciton-polaritons, the manifestation of a coherent interaction between a TMD excitonic transition and a photonic mode. Most of these polaritonic modes have been shown to arise from the strong coupling between two-dimensional (2D) delocalized excitons and bound photonic modes, either obtained with planar microcavities between distributed Bragg reflector (DBR) mirrors \cite{liu_strong_2015, dufferwiel_excitonpolaritons_2015, liu_control_2017} or as guided resonances in planar waveguides \cite{zhang_photonic-crystal_2018, kravtsov_nonlinear_2020, chen_metasurface_2020}. Since these photonic modes are spatially extended in 2D, it would be difficult to realize strong polariton-polariton nonlinearities that are inversely proportional to the confinement area \cite{verger_blockade_2006}, an important requirement for photonic quantum simulators \cite{angelakis_quantum_2017} and strongly correlated photonic devices \cite{gerace_josephson_2009}. While exciton-polaritons in fiber-DBR cavities  \cite{sidler_fermi_2017, gebhardt_polariton_2019} can be confined to a small mode volume, such structures emit light out-of-plane which precludes a straightforward means to couple neighboring cavities. 
	
    On-chip integrated zero-dimensional (0D) cavities, such as photonic crystal defect resonators, provide a means to confine light in a small mode volume while allowing many such cavities to couple to each other via evanescent fields \cite{majumdar_design_2012}. In fact, TMDs coupled to photonic crystal cavities have been used to demonstrate optically pumped lasing \cite{wu_monolayer_2015,li_room-temperature_2017}, cavity enhanced electroluminescence \cite{liu_nanocavity_2017}, and second harmonic generation \cite{fryett_silicon_2016,gan_microwatts_2018}. TMD hetero-structures have also been integrated with photonic crystal resonators to demonstrate emission enhancement \cite{rivera_coupling_2019} and lasing \cite{liu_room_2019}. However, no conclusive signature of a coherent interaction between TMD excitons and a small mode volume nanocavity has been reported.
	
	In this paper we report the signature of coherent coupling between the neutral exciton in monolayer $\text{MoSe}_{2}$ and a 0D photonic crystal nanocavity made of silicon nitride (SiN). Specifically we observed a dispersive shift in the cavity transmission spectrum as the exciton is temperature-tuned near the cavity resonance. Our efforts extend previous attempts to demonstrate coherent coupling between the TMD exciton and a SiN nanobeam cavity \cite{fryett_encapsulated_2018} by implementing a clean dry transfer method \cite{rosser_high-precision_2020} and recognizing the role of exciton-phonon interactions in the cavity coupled photoluminescence \cite{rosser_excitonphonon_2020}. The extracted exciton-cavity coupling from the dispersive shift is $\hbar g \approx 4.3$ \si{\milli\electronvolt} for an estimated cooperativity $C=4g^2/(\kappa_0 \gamma_0) \sim 3.4$, in which $\hbar\gamma_0 = 5.77$ \si{\milli\electronvolt} is the measured intrinsic broadening of the TMD exciton, and $\hbar\kappa_0 = 3.8$ \si{\milli\electronvolt} is the bare photonic mode linewidth measured without the TMD material. While our experiment probes the coupled system in a dispersive regime, it provides a straightforward path to achieve the strong light-matter coupling regime at 4 \si{\kelvin}, with an anticipated cooperativity of $C \sim 380$ for this material platform.
	
	\section*{Results}
	A one-dimensional photonic crystal cavity, also known as a nanobeam resonator, was designed (Fig. \ref{fig:fig1001}a) and fabricated (Fig. \ref{fig:fig1001}b) in a SiN thin-film on a silicon dioxide substrate with an estimated cavity mode volume $V \sim 2 (\lambda / n)^3$, according to the standard cavity QED definition \cite{andreani_strong_1999}. The bare cavity transmission is interrogated using input and output grating couplers. A monolayer $\mathrm{MoSe}_2$ flake was then transferred onto the nanobeam via a modified dry transfer method to eliminate interfering bulk material \cite{rosser_high-precision_2020} (Fig. \ref{fig:fig1001}c, d). This coupled $\mathrm{MoSe}_2$-nanobeam device was placed in a cryostat where the temperature was swept between $80$ \si{\kelvin} and $200$ \si{\kelvin}. 
	
	\subsection*{Device characterization}
	Monolayer $\mathrm{MoSe}_2$ exhibits poor optical contrast on the SiN substrate (Fig. \ref{fig:fig1001}b). Hence, we confirm the presence of the monolayer on the nanocavity by measuring the photoluminescence (PL) (see Materials and Methods). We observed a strong excitonic peak in the PL spectrum, as shown in Fig. \ref{fig:001fig2}a. When the PL is collected from a grating coupler (versus from the full field of view of the confocal microscope), a cavity peak is clearly evidenced in the spectrum  (Fig. \ref{fig:001fig2}b). The background PL is also observed simultaneously due to imperfect spatial filtering in the confocal microscope. Due to a limited field of view in our microscopy setup, the cavity with transferred monolayer must be in close proximity to the output grating, making the spatial filtering of radiation scattered from the sample difficult.
	
	The nanobeam resonator is characterized via resonant transmission. A broadband super-continuum laser is directed into one of the gratings and the transmitted radiation is collected from the other grating. Prior to monolayer material integration, the cavity resonance was measured at 300 \si{\kelvin} to be $\hbar \omega_{C} = 1595$ \si{\milli\electronvolt} with a linewidth $\hbar\kappa_0 = 3.8$ \si{\milli\electronvolt}, corresponding to a bare quality factor $Q_0 = 420$ (Fig. \ref{fig:001fig2}c). After transfer of the monolayer $\mathrm{MoSe}_2$ the bare cavity resonance is measured at 300 \si{\kelvin} to be $\hbar \omega_{C} = 1590$ \si{\milli\electronvolt} with a broadened linewidth $\hbar\kappa = 10.7$ \si{\milli\electronvolt}, corresponding to a loaded quality factor $Q = 149$ (Fig. \ref{fig:001fig2}d). Because the linewidth of the monolayer $\mathrm{MoSe}_2$ is the dominant source of decay the observed quality factor of the device is sufficient to probe the physical effect of dispersive coupling.
	
	\subsection*{Temperature dependence}
	The neutral exciton PL and cavity transmission were then concurrently measured as the temperature was swept from $80$K to $200$K. At low temperature the cavity mode is detuned on the blue side of the excitonic resonance. As the temperature is increased the exciton resonance redshifts, so the detuning between the exciton and cavity resonances decrease. The excitonic PL spectra at different temperatures are fit with a Voigt function \cite{olivero_empirical_1977} to extract the peak energy ($\omega_{X}$) and linewidth ($\gamma$) where we assume the source of inhomogeneous broadening ($\Delta = 4.42 \pm 2.27$ \si{\milli\electronvolt}) is independent of temperature \cite{selig_excitonic_2016}.

	The temperature dependence of the neutral exciton peak energy is fit to the standard equation for the semiconductor bandgap \cite{odonnell_temperature_1991} (Fig. \ref{fig:001fig3}a). This relation corresponds to the neutral exciton energy assuming the exciton binding energy is not strongly temperature dependent. In this experiment the temperature range explored remains in the linear regime at high temperatures.
	\begin{align}\label{energy}
	E_{X} (T) & = E_{X}(0) - S \langle \hbar \omega \rangle [\coth{[ \langle \hbar \omega \rangle / (2 k_{B} T)]} - 1] \nonumber \\
	& \approx E_{X}'(0) - 2 S k_{B} T
	\end{align} where $E_{X}(0)$ is the zero Kelvin neutral exciton energy, $S$ is a dimensionless coupling constant, and $\langle \hbar \omega \rangle$ is the average phonon energy. A fit to the extracted energy of the neutral exciton provides a linearized zero Kelvin neutral exciton energy of $E_{X}'(0) \equiv E_{X}(0) + S \langle \hbar \omega \rangle =  1637$ \si{\milli\electronvolt} and a dimensionless coupling constant $S = 1.21$. These values are comparable to previous reports in the literature \cite{tongay_thermally_2012}. 
	
	Similarly, the temperature dependence of the neutral exciton linewidth is fit to the Rudin equation \cite{rudin_temperature-dependent_1990} (Fig. \ref{fig:001fig3}b).
	
	\begin{align}\label{linewidth}
	\gamma (T) & = \gamma_{0} + c_{1} T + \frac{c_{2}}{e^{\Omega / k_{B} T} - 1} \nonumber \\
	& \approx \gamma_{0}' + R k_{B} T
	\end{align} where $\gamma_{0}$ is the intrinsic homogeneous linewidth, $c_{1}$ includes exciton interactions with acoustic phonons, $c_{2}$ includes exciton interactions with longitudinal-optical phonons, and $\Omega$ is the average phonon energy. In the linearized equation $\gamma_{0}' = \gamma_{0} -\frac{c_{2}}{2}$ and $R = \frac{c_{2}}{\Omega}$ where we have assumed $c_{1} << c_{2}$. A fit to the extracted neutral exciton linewidth provides for an intrinsic linewidth of $\hbar\gamma_{0}' = 5.77$ \si{\milli\electronvolt} and a dimensionless coupling constant $R = 0.69$. These values are also comparable to previous reports in the literature \cite{selig_excitonic_2016}.
	
	It is worth noting the bare nanobeam cavity resonance wavelength does not significantly shift with temperature, which is primarily due to the low thermo-optic coefficient of SiN (Fig. \ref{fig:001fig4}a). However, in the TMD-coupled nanobeam resonator a shift in the cavity resonance is clearly observed as the exciton-cavity detuning decreases (Fig. \ref{fig:001fig4}b). We attribute this shift to the dispersive coupling of the 2D excitons in the monolayer $\mathrm{MoSe}_2$ to the 0D nanobeam cavity mode, which is hereby established via a simple coupled oscillator model.
	
	\subsection*{Coupled Oscillator Model}
	A homogeneous distribution of TMD excitons and a single 0D cavity mode can be phenomenologically modeled with a Hamiltonian describing two coupled oscillators,  wherein the exciton and cavity degrees of freedom coherently interact via an exciton-cavity coupling rate, $g$. The bare oscillator resonance frequencies are measured with respect to a rotating frame at the resonant driving frequency, $\omega_{L}$. The full Hamiltonian is
	\begin{equation}\label{hamiltonian}
	H_{XC} = \hbar \Delta_{XL} a^{\dagger} a + \hbar \Delta_{CL} c^{\dagger} c + \hbar  g ( a^{\dagger} c + c^{\dagger} a ) + E(c + c^{\dagger}) \, ,
	\end{equation} 
	where $\Delta_{XL} = \omega_{X} - \omega_{L}$ and $\Delta_{CL} = \omega_{C} - \omega_{L}$ are the detunings of exciton and cavity mode from the laser frequency, respectively; $a$ ($c$) is the annihilation operator for the exciton (cavity) mode. In the weak excitation regime, exciton saturation and any exciton-exciton interaction can be neglected. Hence, both exciton and cavity operators can be treated as bosonic modes. Including losses, the model can be completed by defining the Liouvillian operator for the density matrix, $\mathcal{L}(\rho) = \frac{1}{i \hbar} [H,\rho] + \hbar\kappa\mathcal{L}_{c}(\rho) + \hbar\gamma\mathcal{L}_{a}(\rho)$, which accounts for the finite cavity and exciton linewidths. The Lindblad operators are $\mathcal{L}_{\xi}(\rho) = \xi \rho \xi^{\dagger} - \frac{1}{2} \xi^{\dagger} \xi \rho - \frac{1}{2} \rho \xi^{\dagger} \xi$, in which $\xi=a,c$.
	
	By diagonalizing the Liouvillian within the single excitation subspace the following eigenenergies can be obtained (Supplementary Note 2) \cite{andreani_strong_1999,hennessy_quantum_2007,englund_controlling_2007,laussy_climbing_2012}
	\begin{equation}\label{eigenenergy}
	\omega_{\pm} = \omega_{C} + \frac{\Delta_{XC}}{2} - i \frac{\kappa + \gamma}{2} \pm \sqrt{g^{2} + \frac{1}{4}\left [ \Delta_{XC} + i (\kappa - \gamma) \right ]^{2}} \, ,
	\end{equation} 
	in which $\Delta_{XC} = \omega_{X} - \omega_{C}$ is the exciton-cavity detuning. A series expansion of the eigenenergies around infinite detuning, $\Delta_{XC} \to \infty$, results in a semi-analytic expression for the cavity peak energy with a far-detuned neutral exciton \cite{blais_cavity_2004}.
	\begin{equation}\label{dispersive}
	\omega_{-} \approx \omega_{C} - \frac{g^{2}}{\Delta_{XC}} \, .
	\end{equation} 
	As a consequence, the cavity resonance is dependent on the exciton-cavity detuning. The experimental data is fit with Eq. \ref{dispersive} for an exciton-cavity coupling energy $\hbar g = 4.27 \pm 0.20$ \si{\milli\electronvolt} (Fig. \ref{fig:001fig4}b). It should be noted the exciton PL peak energy is used as a proxy for the absorption resonance, since $\mathrm{MoSe}_2$ is known to have a small Stokes shift \cite{gebhardt_polariton_2019}. The extracted light-matter interaction is similar to related nanophotonic structures, although we expect it to be larger with an optimal coverage of the cavity mode \cite{rosser_excitonphonon_2020}.
	
\begin{figure}
		\centering
		\includegraphics[width=1.0\linewidth]{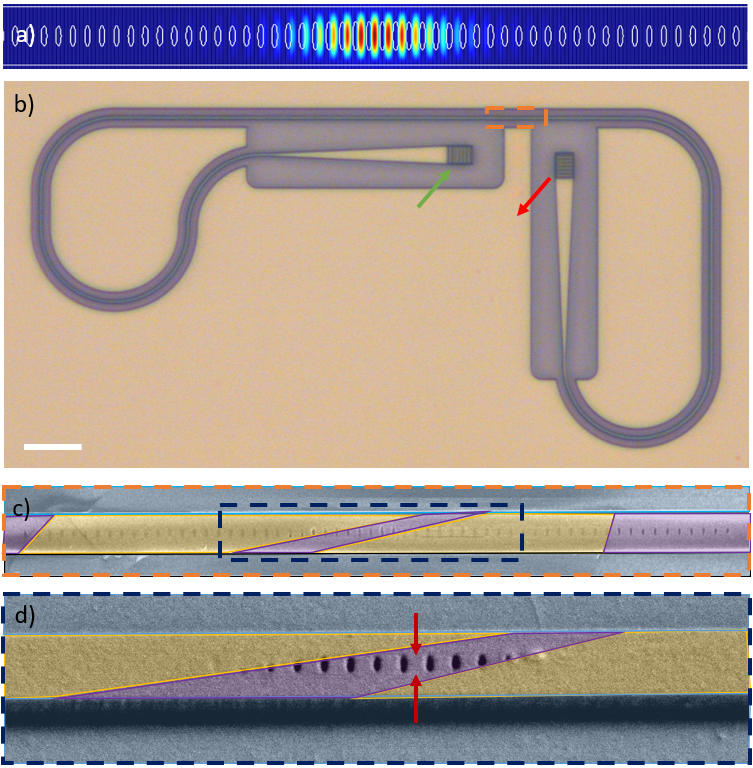}
		\caption{a) Electric field intensity simulated at the center of the SiN nanobeam cavity by 3D-FDTD at the cavity mode resonance frequency, showing wavelength scale field confinement.  b) Optical image of the monolayer $\mathrm{MoSe}_2$ (not visible) integrated onto the nanobeam (orange box) with the grating couplers for transmission measurements (green - excitation, red - collection). Scale bar is $10$ \si{\micro\meter}. c) False color SEM image of the monolayer $\mathrm{MoSe}_2$ integrated onto the nanobeam. ($\mathrm{MoSe}_2$ - gold, SiN - purple, $\mathrm{SiO}_2$ - teal). d) False color SEM image of the monolayer $\mathrm{MoSe}_2$ integrated onto the nanobeam with deposited gold to prevent charging. The obstruction of the nanobeam holes is made explicit. Red arrows indicate the cavity center.}
		\label{fig:fig1001}
	\end{figure}

\begin{figure}
		\centering
		\includegraphics[width=1.0\linewidth]{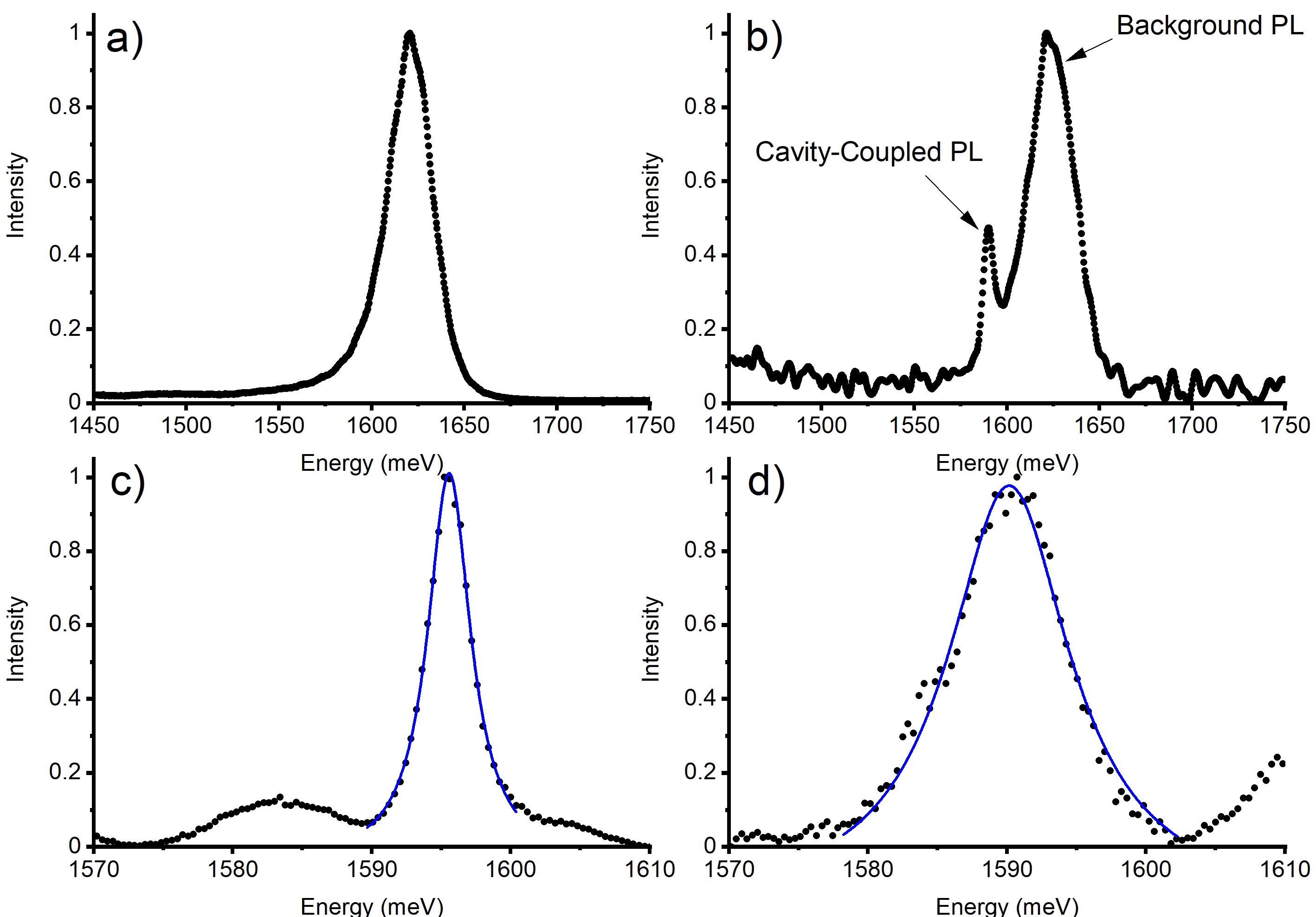}
		\caption{a) Photoluminescence of monolayer $\mathrm{MoSe}_2$ at 80 \si{\kelvin}. b) Cavity-coupled photoluminescence of monolayer $\mathrm{MoSe}_2$ at 80 \si{\kelvin}. Primary peak is background photoluminescence. Secondary peak is collected from the grating coupler confirming cavity coupling. c) Bare transmission spectrum of the nanobeam cavity at 300 \si{\kelvin}. The blue curve is a Lorentzian fit to the cavity resonance. d) Transmission spectrum of the nanobeam cavity with an integrated flake of monolayer $\mathrm{MoSe}_2$ at 300 \si{\kelvin}. The blue curve is a Lorentzian fit to the cavity resonance.}
		\label{fig:001fig2}
	\end{figure}

\begin{figure}
		\centering
		\includegraphics[width=1.0\linewidth]{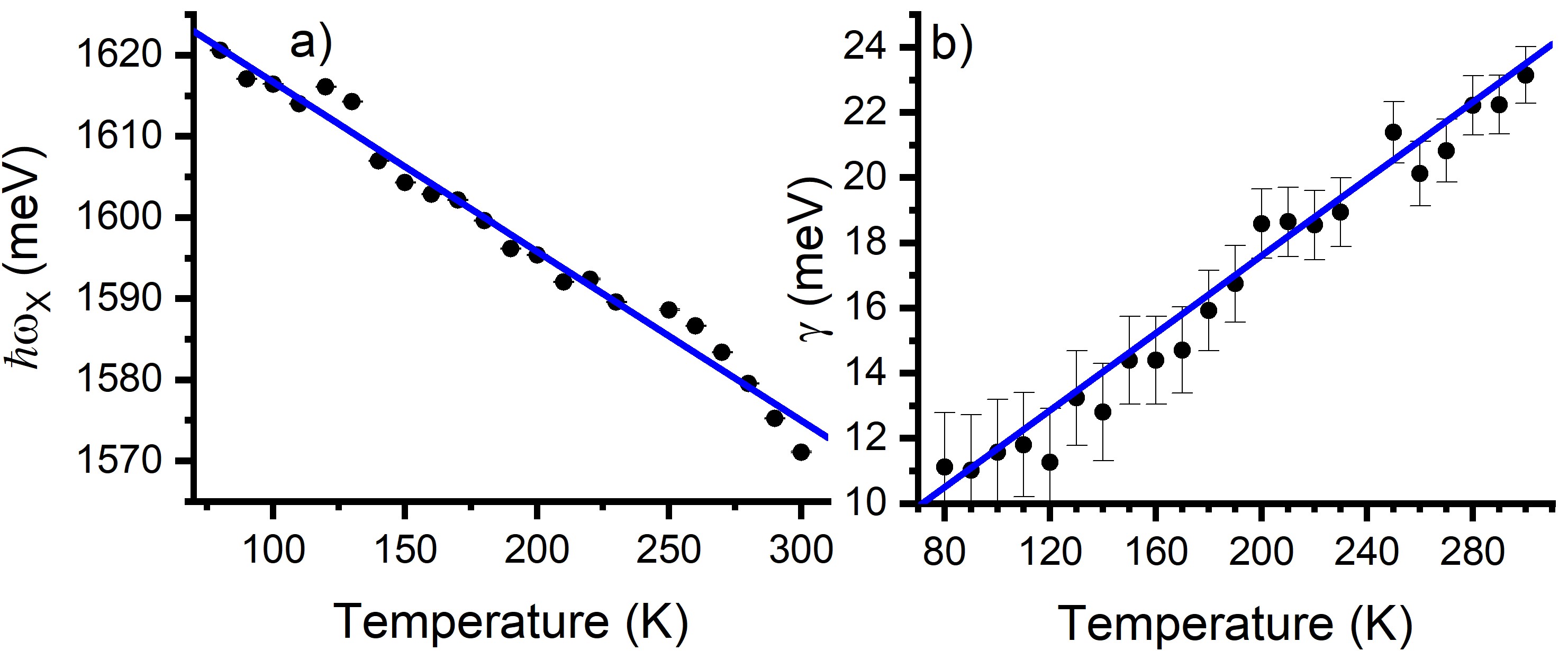}
		\caption{a) Temperature dependence of the neutral exciton resonance.The black dots are the neutral exciton energy observed in the photoluminescence spectrum fit to a Voigt function. The blue line is a fit to Eq. \ref{energy}.  b) Temperature dependence of the neutral exciton linewidth.  The black dots are the neutral exciton linewidth observed in the photoluminescence spectrum fit to a Voigt function. The blue line is a fit to Eq. \ref{linewidth}.}
		\label{fig:001fig3}
	\end{figure}

\begin{figure}
		\centering
		\includegraphics[width=1.0\linewidth]{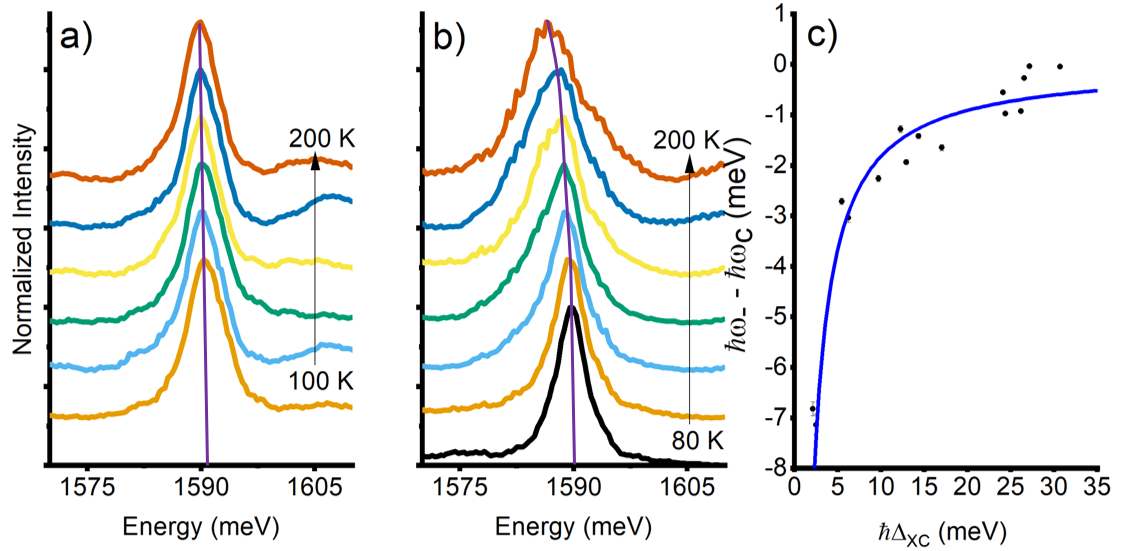}
		\caption{a) Representative transmission spectra of the nanobeam cavity without an integrated flake of monolayer $\mathrm{MoSe}_2$ at 100 \si{\kelvin} to 200 \si{\kelvin} in 20 \si{\kelvin} increments. b) Representative transmission spectra of the nanobeam cavity with an integrated flake of monolayer $\mathrm{MoSe}_2$ at 80 \si{\kelvin} to 200 \si{\kelvin} in 20 \si{\kelvin} increments. b) Dispersive shift of the cavity resonance in transmission.}
		\label{fig:001fig4}
	\end{figure}

	The light-matter coupling energy was numerically simulated, $\hbar g = 4.2$ \si{\milli\electronvolt}, from a theoretical formulation of the exciton dipole interacting with the cavity mode electric field \cite{andreani_polaritons_2014}. We find that $g$ depends on the 2D material extension over the cavity, due to the coupling of a 2D material excitation with a 0D electromagnetic field mode (Supplementary Note 3). Quantitative agreement with the value extracted from the dispersive shift is attained when assuming a 2D flake coverage of the nanocavity compatible with the one inferred from the sample picture (Fig. \ref{fig:fig1001}d). A maximal simulated exciton-cavity coupling energy $\hbar g = 5.1$ \si{\milli\electronvolt} is obtained for this cavity design when the 2D flake extension matches the spatial envelope of the cavity mode electric field. 
	
	\section*{Discussion}
	We have estimated the exciton-cavity coupling of a TMD excitonic transition integrated on a zero-dimensional SiN nanobeam cavity. From the known intrinsic linewidth of the neutral exciton of $\mathrm{MoSe}_2$ encapsulated in boron nitride ($\sim 1.7$ \si{\milli\electronvolt}) \cite{ajayi_approaching_2017}, the strong light-matter coupling regime is within reach at 4 \si{\kelvin}. With an  improvement in the experimental cavity quality factor to $Q = 10000$ \cite{fryett_encapsulated_2018}, and considering that the simulated exciton-cavity mode coupling energy for an optimal 2D material coverage of the cavity surface is $\hbar g = 5.1$ \si{\milli\electronvolt}, a low-temperature cooperativity approaching $C = 4g^{2} / \kappa\gamma \sim 380$ can be anticipated, larger than the state-of-art obtained with single quantum dots \cite{najer_gated_2019}. 
	
	The limiting factor in this material platform is the TMD neutral exciton linewidth, which can be reduced by boron nitride encapsulation \cite{ajayi_approaching_2017}. However, there is a trade-off with a reduced field overlap of the cavity mode with the monolayer material due to the increased physical distance from the cavity field maximum. This consequence can be ameliorated by using an air mode cavity \cite{quan_deterministic_2011}. The primary concern with optimization of the cavity design and bare quality factor (without vdW materials) may be design tolerance with respect to cavity perturbations.
	
	An alternative means of integrating the TMD material on the nanocavity without the precision method used here could include large-area transfers using CVD grown monolayers or template stripping methods \cite{chen_chemical_2017, liu_disassembling_2020}. Subsequent etching would allow for precise control of the geometry of the vdW material on the nanocavity. This would allow for studies in cavity perturbation theory \cite{yang_simple_2015} as well as careful elucidation of the non-linearities inherent to the monolayer neutral exciton \cite{shahnazaryan_exciton-exciton_2017}. A large-area transfer method would allow for digital tuning of the cavity resonance using an ensemble of cavities instead of tuning the exciton energy via temperature.
	
	\section*{Materials and Methods}
	
	\subsection*{Design and Fabrication}
	
	The nanobeam is made of a $t=220$ \si{\nano\meter} thick and a $w = 779$ \si{\nano\meter} wide silicon nitride film on silicon oxide substrate. The center region of the nanobeam, where the light is confined, consists of 10 tapering elliptical holes, and the reflectors are made of 20 Bragg mirror holes. The minor axis radius of the elliptical holes is fixed to $40$ \si{\nano\meter}. The tapering region begins with a $178$ \si{\nano\meter} major axis diameter and a $215$ \si{\nano\meter} center-to-center distance. The tapering region is quadratically tapered to a $121$ \si{\nano\meter} major axis radius and a $233$ \si{\nano\meter} center-to-center distance. The Bragg region remains with fixed values of a $121$ \si{\nano\meter} major axis radius and a $233$ \si{\nano\meter} center-to-center distance. The performance of the nanobeam cavity is optimized using Lumerical FDTD solutions software. With these design parameters, we obtain a theoretical resonance at 1639 meV and an intrinsic theoretical Q=25000.
		
	The cavity is fabricated using a 220 \si{\nano \meter} thick SiN membrane grown via LPCVD on 4 \si{\micro \meter} of thermal oxide on silicon. The samples were obtained from commercial vendor Rogue Valley Microelectronics. A piece of the wafer is spin coated with roughly 400 \si{\nano \meter} of Zeon ZEP520A resist. The resist is further coated with a thin layer of Pt/Au which serves as a charge dissipation layer, as both SiN and silicon dioxide are insulators. The resist was then patterned using a JEOL JBX6300FX electron-beam lithography system with an accelerating voltage of 100 \si{\kilo \volt}. The pattern was transferred to the SiN using a reactive ion etch (RIE) in $\text{CHF}_{3} \text{/O}_{2}$ chemistry.
	The fabricated sample shows a blue shift of the cavity resonance as compared to the simulated value, as well as significant reduction in the measured Q-factor, which is attributed to fabrication imperfections.

	\subsection*{Measurement}
	
	The photoluminescence spectrum is measured by exciting the monolayer with a 632 \si{\nano\meter} Helium-Neon laser. The resulting emission is collected with a free-space confocal microscopy setup and directed into a Princeton Instruments IsoPlane SCT-320 Imaging Spectrograph. Cavity-coupled photoluminescence spectrum is measured by exciting the monolayer with the same 632 \si{\nano\meter} Helium-Neon laser and collecting from a grating coupler with the direct photoluminescence emission occluded by a pinhole in the image plane of the confocal microscope. The transmission spectrum is measured by exciting one of the grating couplers with a supercontinuum laser (Fianium WhiteLase Micro) and collecting from the other grating coupler. Using liquid nitrogen in a continuous flow cryostat (Janis ST-500) the energy of the neutral exciton in the monolayer $\mathrm{MoSe}_2$ is shifted with consequent changes in the linewidths.
	
	\section*{Acknowledgments}
	The research was supported by NSF-1845009 and NSF-ECCS-1708579. D.R. is partially supported by a CEI graduate fellowship. A.R. acknowledges support from the IC Postdoctoral Research Fellowship. Part of this work was conducted at the Washington Nanofabrication Facility / Molecular Analysis Facility, a National Nanotechnology Coordinated Infrastructure (NNCI) site at the University of Washington, which is supported in part by funds from the National Science Foundation (awards NNCI-1542101, 1337840 and 0335765), the National Institutes of Health, the Molecular Engineering \& Sciences Institute, the Clean Energy Institute, the Washington Research Foundation, the M. J. Murdock Charitable Trust, Altatech, ClassOne Technology, GCE Market, Google and SPTS.

\bibliography{references}

\begin{thebibliography}{41}
\providecommand{\natexlab}[1]{#1}
\providecommand{\url}[1]{\texttt{#1}}
\expandafter\ifx\csname urlstyle\endcsname\relax
  \providecommand{\doi}[1]{doi: #1}\else
  \providecommand{\doi}{doi: \begingroup \urlstyle{rm}\Url}\fi

\bibitem[Liu et~al.(2019{\natexlab{a}})Liu, Zheng, Chen, Fryett, and
  Majumdar]{liu_van_2019}
Chang-hua Liu, Jiajiu Zheng, Yueyang Chen, Taylor Fryett, and Arka Majumdar.
\newblock Van der {Waals} materials integrated nanophotonic devices
  [{Invited}].
\newblock \emph{Optical Materials Express}, 9\penalty0 (2):\penalty0 384--399,
  February 2019{\natexlab{a}}.
\newblock ISSN 2159-3930.
\newblock \doi{10.1364/OME.9.000384}.
\newblock URL
  \url{https://www.osapublishing.org/ome/abstract.cfm?uri=ome-9-2-384}.
\newblock Publisher: Optical Society of America.

\bibitem[Liu et~al.(2015)Liu, Galfsky, Sun, Xia, Lin, Lee, K{\'e}na-Cohen, and
  Menon]{liu_strong_2015}
Xiaoze Liu, Tal Galfsky, Zheng Sun, Fengnian Xia, Erh-chen Lin, Yi-Hsien Lee,
  St{\'e}phane K{\'e}na-Cohen, and Vinod~M. Menon.
\newblock Strong light{\textendash}matter coupling in two-dimensional atomic
  crystals.
\newblock \emph{Nature Photonics}, 9\penalty0 (1):\penalty0 30--34, January
  2015.
\newblock ISSN 1749-4893.
\newblock \doi{10.1038/nphoton.2014.304}.
\newblock URL \url{https://www.nature.com/articles/nphoton.2014.304}.
\newblock Number: 1 Publisher: Nature Publishing Group.

\bibitem[Dufferwiel et~al.(2015)Dufferwiel, Schwarz, Withers, Trichet, Li,
  Sich, Del Pozo-Zamudio, Clark, Nalitov, Solnyshkov, Malpuech, Novoselov,
  Smith, Skolnick, Krizhanovskii, and
  Tartakovskii]{dufferwiel_excitonpolaritons_2015}
S.~Dufferwiel, S.~Schwarz, F.~Withers, A.~a.~P. Trichet, F.~Li, M.~Sich, O.~Del
  Pozo-Zamudio, C.~Clark, A.~Nalitov, D.~D. Solnyshkov, G.~Malpuech, K.~S.
  Novoselov, J.~M. Smith, M.~S. Skolnick, D.~N. Krizhanovskii, and A.~I.
  Tartakovskii.
\newblock Exciton{\textendash}polaritons in van der {Waals} heterostructures
  embedded in tunable microcavities.
\newblock \emph{Nature Communications}, 6\penalty0 (1):\penalty0 8579, October
  2015.
\newblock ISSN 2041-1723.
\newblock \doi{10.1038/ncomms9579}.
\newblock URL \url{https://www.nature.com/articles/ncomms9579}.
\newblock Number: 1 Publisher: Nature Publishing Group.

\bibitem[Liu et~al.(2017{\natexlab{a}})Liu, Bao, Li, Ropp, Wang, and
  Zhang]{liu_control_2017}
Xiaoze Liu, Wei Bao, Quanwei Li, Chad Ropp, Yuan Wang, and Xiang Zhang.
\newblock Control of {Coherently} {Coupled} {Exciton} {Polaritons} in
  {Monolayer} {Tungsten} {Disulphide}.
\newblock \emph{Physical Review Letters}, 119\penalty0 (2):\penalty0 027403,
  July 2017{\natexlab{a}}.
\newblock \doi{10.1103/PhysRevLett.119.027403}.
\newblock URL \url{https://link.aps.org/doi/10.1103/PhysRevLett.119.027403}.
\newblock Publisher: American Physical Society.

\bibitem[Zhang et~al.(2018)Zhang, Gogna, Burg, Tutuc, and
  Deng]{zhang_photonic-crystal_2018}
Long Zhang, Rahul Gogna, Will Burg, Emanuel Tutuc, and Hui Deng.
\newblock Photonic-crystal exciton-polaritons in monolayer semiconductors.
\newblock \emph{Nature Communications}, 9\penalty0 (1):\penalty0 713, February
  2018.
\newblock ISSN 2041-1723.
\newblock \doi{10.1038/s41467-018-03188-x}.
\newblock URL \url{https://www.nature.com/articles/s41467-018-03188-x}.
\newblock Number: 1 Publisher: Nature Publishing Group.

\bibitem[Kravtsov et~al.(2020)Kravtsov, Khestanova, Benimetskiy, Ivanova,
  Samusev, Sinev, Pidgayko, Mozharov, Mukhin, Lozhkin, Kapitonov, Brichkin,
  Kulakovskii, Shelykh, Tartakovskii, Walker, Skolnick, Krizhanovskii, and
  Iorsh]{kravtsov_nonlinear_2020}
Vasily Kravtsov, Ekaterina Khestanova, Fedor~A. Benimetskiy, Tatiana Ivanova,
  Anton~K. Samusev, Ivan~S. Sinev, Dmitry Pidgayko, Alexey~M. Mozharov, Ivan~S.
  Mukhin, Maksim~S. Lozhkin, Yuri~V. Kapitonov, Andrey~S. Brichkin, Vladimir~D.
  Kulakovskii, Ivan~A. Shelykh, Alexander~I. Tartakovskii, Paul~M. Walker,
  Maurice~S. Skolnick, Dmitry~N. Krizhanovskii, and Ivan~V. Iorsh.
\newblock Nonlinear polaritons in a monolayer semiconductor coupled to optical
  bound states in the continuum.
\newblock \emph{Light: Science \& Applications}, 9\penalty0 (1):\penalty0 56,
  April 2020.
\newblock ISSN 2047-7538.
\newblock \doi{10.1038/s41377-020-0286-z}.
\newblock URL \url{https://www.nature.com/articles/s41377-020-0286-z}.
\newblock Number: 1 Publisher: Nature Publishing Group.

\bibitem[Chen et~al.(2020)Chen, Miao, Wang, Zhong, Saxena, Chow, Whitehead,
  Gerace, Xu, Shi, and Majumdar]{chen_metasurface_2020}
Yueyang Chen, Shengnan Miao, Tianmeng Wang, Ding Zhong, Abhi Saxena, Colin
  Chow, James Whitehead, Dario Gerace, Xiaodong Xu, Su-Fei Shi, and Arka
  Majumdar.
\newblock Metasurface {Integrated} {Monolayer} {Exciton} {Polariton}.
\newblock \emph{Nano Letters}, June 2020.
\newblock ISSN 1530-6984.
\newblock \doi{10.1021/acs.nanolett.0c01624}.
\newblock URL \url{https://doi.org/10.1021/acs.nanolett.0c01624}.
\newblock Publisher: American Chemical Society.

\bibitem[Verger et~al.(2006)Verger, Ciuti, and Carusotto]{verger_blockade_2006}
A.~Verger, C.~Ciuti, and I.~Carusotto.
\newblock Polariton quantum blockade in a photonic dot.
\newblock \emph{Phys. Rev. B}, 73:\penalty0 193306, May 2006.
\newblock \doi{10.1103/PhysRevB.73.193306}.
\newblock URL \url{https://link.aps.org/doi/10.1103/PhysRevB.73.193306}.

\bibitem[Angelakis(2017)]{angelakis_quantum_2017}
Dimitris~G. Angelakis, editor.
\newblock \emph{Quantum {Simulations} with {Photons} and {Polaritons}:
  {Merging} {Quantum} {Optics} with {Condensed} {Matter} {Physics}}.
\newblock Quantum {Science} and {Technology}. Springer International
  Publishing, 2017.
\newblock ISBN 978-3-319-52023-0.
\newblock \doi{10.1007/978-3-319-52025-4}.
\newblock URL \url{https://www.springer.com/gp/book/9783319520230}.

\bibitem[Gerace et~al.(2009)Gerace, T{\"u}reci, Imamoglu, Giovannetti, and
  Fazio]{gerace_josephson_2009}
Dario Gerace, Hakan~E. T{\"u}reci, Atac Imamoglu, Vittorio Giovannetti, and
  Rosario Fazio.
\newblock The quantum-optical josephson interferometer.
\newblock \emph{Nature Physics}, 5\penalty0 (4):\penalty0 281--284, 2009.
\newblock \doi{10.1038/nphys1223}.
\newblock URL \url{https://doi.org/10.1038/nphys1223}.

\bibitem[Sidler et~al.(2017)Sidler, Back, Cotlet, Srivastava, Fink, Kroner,
  Demler, and Imamoglu]{sidler_fermi_2017}
Meinrad Sidler, Patrick Back, Ovidiu Cotlet, Ajit Srivastava, Thomas Fink,
  Martin Kroner, Eugene Demler, and Atac Imamoglu.
\newblock Fermi polaron-polaritons in charge-tunable atomically thin
  semiconductors.
\newblock \emph{Nature Physics}, 13\penalty0 (3):\penalty0 255--261, March
  2017.
\newblock ISSN 1745-2481.
\newblock \doi{10.1038/nphys3949}.
\newblock URL \url{https://www.nature.com/articles/nphys3949}.
\newblock Number: 3 Publisher: Nature Publishing Group.

\bibitem[Gebhardt et~al.(2019)Gebhardt, F{\"o}rg, Yamaguchi, Bilgin, Mohite,
  Gies, Florian, Hartmann, H{\"a}nsch, H{\"o}gele, and
  Hunger]{gebhardt_polariton_2019}
Christian Gebhardt, Michael F{\"o}rg, Hisato Yamaguchi, Ismail Bilgin,
  Aditya~D. Mohite, Christopher Gies, Matthias Florian, Malte Hartmann,
  Theodor~W. H{\"a}nsch, Alexander H{\"o}gele, and David Hunger.
\newblock Polariton hyperspectral imaging of two-dimensional semiconductor
  crystals.
\newblock \emph{Scientific Reports}, 9\penalty0 (1):\penalty0 13756, September
  2019.
\newblock ISSN 2045-2322.
\newblock \doi{10.1038/s41598-019-50316-8}.
\newblock URL \url{https://www.nature.com/articles/s41598-019-50316-8}.
\newblock Number: 1 Publisher: Nature Publishing Group.

\bibitem[Majumdar et~al.(2012)Majumdar, Rundquist, Bajcsy, Dasika, Bank, and
  Vu{\v c}kovi{\'c}]{majumdar_design_2012}
Arka Majumdar, Armand Rundquist, Michal Bajcsy, Vaishno~D. Dasika, Seth~R.
  Bank, and Jelena Vu{\v c}kovi{\'c}.
\newblock Design and analysis of photonic crystal coupled cavity arrays for
  quantum simulation.
\newblock \emph{Physical Review B}, 86\penalty0 (19):\penalty0 195312, November
  2012.
\newblock \doi{10.1103/PhysRevB.86.195312}.
\newblock URL \url{https://link.aps.org/doi/10.1103/PhysRevB.86.195312}.
\newblock Publisher: American Physical Society.

\bibitem[Wu et~al.(2015)Wu, Buckley, Schaibley, Feng, Yan, Mandrus, Hatami,
  Yao, Vu{\v c}kovi{\'c}, Majumdar, and Xu]{wu_monolayer_2015}
Sanfeng Wu, Sonia Buckley, John~R. Schaibley, Liefeng Feng, Jiaqiang Yan,
  David~G. Mandrus, Fariba Hatami, Wang Yao, Jelena Vu{\v c}kovi{\'c}, Arka
  Majumdar, and Xiaodong Xu.
\newblock Monolayer semiconductor nanocavity lasers with ultralow thresholds.
\newblock \emph{Nature}, 520\penalty0 (7545):\penalty0 69--72, April 2015.
\newblock ISSN 1476-4687.
\newblock \doi{10.1038/nature14290}.
\newblock URL \url{https://www.nature.com/articles/nature14290}.
\newblock Number: 7545 Publisher: Nature Publishing Group.

\bibitem[Li et~al.(2017)Li, Zhang, Huang, Sun, Fan, Feng, Wang, and
  Ning]{li_room-temperature_2017}
Yongzhuo Li, Jianxing Zhang, Dandan Huang, Hao Sun, Fan Fan, Jiabin Feng, Zhen
  Wang, and C.~Z. Ning.
\newblock Room-temperature continuous-wave lasing from monolayer molybdenum
  ditelluride integrated with a silicon nanobeam cavity.
\newblock \emph{Nature Nanotechnology}, 12\penalty0 (10):\penalty0 987--992,
  October 2017.
\newblock ISSN 1748-3395.
\newblock \doi{10.1038/nnano.2017.128}.
\newblock URL \url{https://www.nature.com/articles/nnano.2017.128}.
\newblock Number: 10 Publisher: Nature Publishing Group.

\bibitem[Liu et~al.(2017{\natexlab{b}})Liu, Clark, Fryett, Wu, Zheng, Hatami,
  Xu, and Majumdar]{liu_nanocavity_2017}
Chang-Hua Liu, Genevieve Clark, Taylor Fryett, Sanfeng Wu, Jiajiu Zheng, Fariba
  Hatami, Xiaodong Xu, and Arka Majumdar.
\newblock Nanocavity {Integrated} van der {Waals} {Heterostructure}
  {Light}-{Emitting} {Tunneling} {Diode}.
\newblock \emph{Nano Letters}, 17\penalty0 (1):\penalty0 200--205, January
  2017{\natexlab{b}}.
\newblock ISSN 1530-6984.
\newblock \doi{10.1021/acs.nanolett.6b03801}.
\newblock URL \url{https://doi.org/10.1021/acs.nanolett.6b03801}.
\newblock Publisher: American Chemical Society.

\bibitem[Fryett et~al.(2016)Fryett, Seyler, Zheng, Liu, Xu, and
  Majumdar]{fryett_silicon_2016}
Taylor~K. Fryett, Kyle~L. Seyler, Jiajiu Zheng, Chang-Hua Liu, Xiaodong Xu, and
  Arka Majumdar.
\newblock Silicon photonic crystal cavity enhanced second-harmonic generation
  from monolayer {WSe} 2.
\newblock \emph{2D Materials}, 4\penalty0 (1):\penalty0 015031, December 2016.
\newblock ISSN 2053-1583.
\newblock \doi{10.1088/2053-1583/4/1/015031}.
\newblock URL \url{https://doi.org/10.1088%2F2053-1583%2F4%2F1%2F015031}.
\newblock Publisher: IOP Publishing.

\bibitem[Gan et~al.(2018)Gan, Zhao, Hu, Wang, Song, Li, Zhao, Jie, and
  Zhao]{gan_microwatts_2018}
Xue-Tao Gan, Chen-Yang Zhao, Si-Qi Hu, Tao Wang, Yu~Song, Jie Li, Qing-Hua
  Zhao, Wan-Qi Jie, and Jian-Lin Zhao.
\newblock Microwatts continuous-wave pumped second harmonic generation in few-
  and mono-layer {GaSe}.
\newblock \emph{Light: Science \& Applications}, 7\penalty0 (1):\penalty0
  17126--17126, January 2018.
\newblock ISSN 2047-7538.
\newblock \doi{10.1038/lsa.2017.126}.
\newblock URL \url{https://www.nature.com/articles/lsa2017126}.
\newblock Number: 1 Publisher: Nature Publishing Group.

\bibitem[Rivera et~al.(2019)Rivera, Fryett, Chen, Liu, Ray, Hatami, Yan,
  Mandrus, Yao, Majumdar, and Xu]{rivera_coupling_2019}
Pasqual Rivera, Taylor~K. Fryett, Yueyang Chen, Chang-Hua Liu, Essance Ray,
  Fariba Hatami, Jiaqiang Yan, David Mandrus, Wang Yao, Arka Majumdar, and
  Xiaodong Xu.
\newblock Coupling of photonic crystal cavity and interlayer exciton in
  heterobilayer of transition metal dichalcogenides.
\newblock \emph{2D Materials}, 7\penalty0 (1):\penalty0 015027, December 2019.
\newblock ISSN 2053-1583.
\newblock \doi{10.1088/2053-1583/ab597d}.
\newblock URL \url{https://doi.org/10.1088%2F2053-1583%2Fab597d}.
\newblock Publisher: IOP Publishing.

\bibitem[Liu et~al.(2019{\natexlab{b}})Liu, Fang, Rasmita, Zhou, Li, Yu, Xiong,
  Zheludev, Liu, and Gao]{liu_room_2019}
Yuanda Liu, Hanlin Fang, Abdullah Rasmita, Yu~Zhou, Juntao Li, Ting Yu, Qihua
  Xiong, Nikolay Zheludev, Jin Liu, and Weibo Gao.
\newblock Room temperature nanocavity laser with interlayer excitons in {2D}
  heterostructures.
\newblock \emph{Science Advances}, 5\penalty0 (4):\penalty0 eaav4506, April
  2019{\natexlab{b}}.
\newblock ISSN 2375-2548.
\newblock \doi{10.1126/sciadv.aav4506}.
\newblock URL \url{https://advances.sciencemag.org/content/5/4/eaav4506}.
\newblock Publisher: American Association for the Advancement of Science
  Section: Research Article.

\bibitem[Fryett et~al.(2018)Fryett, Chen, Whitehead, Peycke, Xu, and
  Majumdar]{fryett_encapsulated_2018}
Taylor~K. Fryett, Yueyang Chen, James Whitehead, Zane~Matthew Peycke, Xiaodong
  Xu, and Arka Majumdar.
\newblock Encapsulated {Silicon} {Nitride} {Nanobeam} {Cavity} for {Hybrid}
  {Nanophotonics}.
\newblock \emph{ACS Photonics}, 5\penalty0 (6):\penalty0 2176--2181, June 2018.
\newblock \doi{10.1021/acsphotonics.8b00036}.
\newblock URL \url{https://doi.org/10.1021/acsphotonics.8b00036}.
\newblock Publisher: American Chemical Society.

\bibitem[Rosser et~al.(2020{\natexlab{a}})Rosser, Fryett, Saxena, Ryou,
  Majumdar, and Majumdar]{rosser_high-precision_2020}
David Rosser, Taylor Fryett, Abhi Saxena, Albert Ryou, Arka Majumdar, and Arka
  Majumdar.
\newblock High-precision local transfer of van der {Waals} materials on
  nanophotonic structures.
\newblock \emph{Optical Materials Express}, 10\penalty0 (2):\penalty0 645--652,
  February 2020{\natexlab{a}}.
\newblock ISSN 2159-3930.
\newblock \doi{10.1364/OME.383255}.
\newblock URL
  \url{https://www.osapublishing.org/ome/abstract.cfm?uri=ome-10-2-645}.
\newblock Publisher: Optical Society of America.

\bibitem[Rosser et~al.(2020{\natexlab{b}})Rosser, Fryett, Ryou, Saxena, and
  Majumdar]{rosser_excitonphonon_2020}
David Rosser, Taylor Fryett, Albert Ryou, Abhi Saxena, and Arka Majumdar.
\newblock Exciton{\textendash}phonon interactions in nanocavity-integrated
  monolayer transition metal dichalcogenides.
\newblock \emph{npj 2D Materials and Applications}, 4\penalty0 (1):\penalty0
  1--6, July 2020{\natexlab{b}}.
\newblock ISSN 2397-7132.
\newblock \doi{10.1038/s41699-020-0156-9}.
\newblock URL \url{https://www.nature.com/articles/s41699-020-0156-9}.
\newblock Number: 1 Publisher: Nature Publishing Group.

\bibitem[Andreani et~al.(1999)Andreani, Panzarini, and
  G\'erard]{andreani_strong_1999}
Lucio~Claudio Andreani, Giovanna Panzarini, and Jean-Michel G\'erard.
\newblock Strong-coupling regime for quantum boxes in pillar microcavities:
  Theory.
\newblock \emph{Phys. Rev. B}, 60:\penalty0 13276--13279, Nov 1999.
\newblock \doi{10.1103/PhysRevB.60.13276}.
\newblock URL \url{https://link.aps.org/doi/10.1103/PhysRevB.60.13276}.

\bibitem[Olivero and Longbothum(1977)]{olivero_empirical_1977}
J.~J. Olivero and R.~L. Longbothum.
\newblock Empirical fits to the {Voigt} line width: {A} brief review.
\newblock \emph{Journal of Quantitative Spectroscopy and Radiative Transfer},
  17\penalty0 (2):\penalty0 233--236, February 1977.
\newblock ISSN 0022-4073.
\newblock \doi{10.1016/0022-4073(77)90161-3}.
\newblock URL
  \url{http://www.sciencedirect.com/science/article/pii/0022407377901613}.

\bibitem[Selig et~al.(2016)Selig, Bergh{\"a}user, Raja, Nagler, Sch{\"u}ller,
  Heinz, Korn, Chernikov, Malic, and Knorr]{selig_excitonic_2016}
Malte Selig, Gunnar Bergh{\"a}user, Archana Raja, Philipp Nagler, Christian
  Sch{\"u}ller, Tony~F. Heinz, Tobias Korn, Alexey Chernikov, Ermin Malic, and
  Andreas Knorr.
\newblock Excitonic linewidth and coherence lifetime in monolayer transition
  metal dichalcogenides.
\newblock \emph{Nature Communications}, 7\penalty0 (1):\penalty0 13279,
  November 2016.
\newblock ISSN 2041-1723.
\newblock \doi{10.1038/ncomms13279}.
\newblock URL \url{https://www.nature.com/articles/ncomms13279}.
\newblock Number: 1 Publisher: Nature Publishing Group.

\bibitem[O{\textquoteright}Donnell and Chen(1991)]{odonnell_temperature_1991}
K.~P. O{\textquoteright}Donnell and X.~Chen.
\newblock Temperature dependence of semiconductor band gaps.
\newblock \emph{Applied Physics Letters}, 58\penalty0 (25):\penalty0
  2924--2926, June 1991.
\newblock ISSN 0003-6951.
\newblock \doi{10.1063/1.104723}.
\newblock URL \url{https://aip.scitation.org/doi/10.1063/1.104723}.
\newblock Publisher: American Institute of Physics.

\bibitem[Tongay et~al.(2012)Tongay, Zhou, Ataca, Lo, Matthews, Li, Grossman,
  and Wu]{tongay_thermally_2012}
Sefaattin Tongay, Jian Zhou, Can Ataca, Kelvin Lo, Tyler~S. Matthews, Jingbo
  Li, Jeffrey~C. Grossman, and Junqiao Wu.
\newblock Thermally {Driven} {Crossover} from {Indirect} toward {Direct}
  {Bandgap} in {2D} {Semiconductors}: {MoSe2} versus {MoS2}.
\newblock \emph{Nano Letters}, 12\penalty0 (11):\penalty0 5576--5580, November
  2012.
\newblock ISSN 1530-6984.
\newblock \doi{10.1021/nl302584w}.
\newblock URL \url{https://doi.org/10.1021/nl302584w}.
\newblock Publisher: American Chemical Society.

\bibitem[Rudin and Reinecke(1990)]{rudin_temperature-dependent_1990}
S.~Rudin and T.~L. Reinecke.
\newblock Temperature-dependent exciton linewidths in semiconductor quantum
  wells.
\newblock \emph{Physical Review B}, 41\penalty0 (5):\penalty0 3017--3027,
  February 1990.
\newblock \doi{10.1103/PhysRevB.41.3017}.
\newblock URL \url{https://link.aps.org/doi/10.1103/PhysRevB.41.3017}.
\newblock Publisher: American Physical Society.

\bibitem[Hennessy et~al.(2007)Hennessy, Badolato, Winger, Gerace, Atat{\"u}re,
  Gulde, F{\"a}lt, Hu, and Imamo{\u g}lu]{hennessy_quantum_2007}
K.~Hennessy, A.~Badolato, M.~Winger, D.~Gerace, M.~Atat{\"u}re, S.~Gulde,
  S.~F{\"a}lt, E.~L. Hu, and A.~Imamo{\u g}lu.
\newblock Quantum nature of a strongly coupled single quantum dot--cavity
  system.
\newblock \emph{Nature}, 445\penalty0 (7130):\penalty0 896--899, 2007.
\newblock \doi{10.1038/nature05586}.
\newblock URL \url{https://doi.org/10.1038/nature05586}.

\bibitem[Englund et~al.(2007)Englund, Faraon, Fushman, Stoltz, Petroff, and
  Vu{\v c}kovi{\'c}]{englund_controlling_2007}
Dirk Englund, Andrei Faraon, Ilya Fushman, Nick Stoltz, Pierre Petroff, and
  Jelena Vu{\v c}kovi{\'c}.
\newblock Controlling cavity reflectivity with a single quantum dot.
\newblock \emph{Nature}, 450\penalty0 (7171):\penalty0 857--861, December 2007.
\newblock ISSN 1476-4687.
\newblock \doi{10.1038/nature06234}.
\newblock URL \url{https://www.nature.com/articles/nature06234}.
\newblock Number: 7171 Publisher: Nature Publishing Group.

\bibitem[Laussy et~al.(2012)Laussy, Valle, Schrapp, Laucht, and
  Finley]{laussy_climbing_2012}
Fabrice~P. Laussy, Elena~del Valle, Michael Schrapp, Arne Laucht, and
  Jonathan~J. Finley.
\newblock Climbing the {Jaynes}-{Cummings} ladder by photon counting.
\newblock \emph{Journal of Nanophotonics}, 6\penalty0 (1):\penalty0 061803,
  November 2012.
\newblock ISSN 1934-2608, 1934-2608.
\newblock \doi{10.1117/1.JNP.6.061803}.
\newblock URL
  \url{https://www.spiedigitallibrary.org/journals/Journal-of-Nanophotonics/volume-6/issue-1/061803/Climbing-the-Jaynes-Cummings-ladder-by-photon-counting/10.1117/1.JNP.6.061803.short}.
\newblock Publisher: International Society for Optics and Photonics.

\bibitem[Blais et~al.(2004)Blais, Huang, Wallraff, Girvin, and
  Schoelkopf]{blais_cavity_2004}
Alexandre Blais, Ren-Shou Huang, Andreas Wallraff, S.~M. Girvin, and R.~J.
  Schoelkopf.
\newblock Cavity quantum electrodynamics for superconducting electrical
  circuits: {An} architecture for quantum computation.
\newblock \emph{Physical Review A}, 69\penalty0 (6):\penalty0 062320, June
  2004.
\newblock \doi{10.1103/PhysRevA.69.062320}.
\newblock URL \url{https://link.aps.org/doi/10.1103/PhysRevA.69.062320}.
\newblock Publisher: American Physical Society.

\bibitem[Andreani(2014)]{andreani_polaritons_2014}
Lucio~Claudio Andreani.
\newblock \emph{in ``Strong Light-Matter Coupling: from atoms to solid state
  systems''}, chapter~2, pages 37--82.
\newblock World Scientific, Singapore, 2014.
\newblock \doi{10.1142/9789814460354_0002}.
\newblock URL
  \url{https://www.worldscientific.com/doi/abs/10.1142/9789814460354_0002}.

\bibitem[Ajayi et~al.(2017)Ajayi, Ardelean, Shepard, Wang, Antony, Taniguchi,
  Watanabe, Heinz, Strauf, Zhu, and Hone]{ajayi_approaching_2017}
Obafunso~A. Ajayi, Jenny~V. Ardelean, Gabriella~D. Shepard, Jue Wang,
  Abhinandan Antony, Takeshi Taniguchi, Kenji Watanabe, Tony~F. Heinz, Stefan
  Strauf, X.-Y. Zhu, and James~C. Hone.
\newblock Approaching the intrinsic photoluminescence linewidth in transition
  metal dichalcogenide monolayers.
\newblock \emph{2D Materials}, 4\penalty0 (3):\penalty0 031011, July 2017.
\newblock ISSN 2053-1583.
\newblock \doi{10.1088/2053-1583/aa6aa1}.
\newblock URL \url{https://doi.org/10.1088%2F2053-1583%2Faa6aa1}.
\newblock Publisher: IOP Publishing.

\bibitem[Najer et~al.(2019)Najer, S{\"o}llner, Sekatski, Dolique, L{\"o}bl,
  Riedel, Schott, Starosielec, Valentin, Wieck, Sangouard, Ludwig, and
  Warburton]{najer_gated_2019}
Daniel Najer, Immo S{\"o}llner, Pavel Sekatski, Vincent Dolique, Matthias~C.
  L{\"o}bl, Daniel Riedel, R{\"u}diger Schott, Sebastian Starosielec, Sascha~R.
  Valentin, Andreas~D. Wieck, Nicolas Sangouard, Arne Ludwig, and Richard~J.
  Warburton.
\newblock A gated quantum dot strongly coupled to an optical microcavity.
\newblock \emph{Nature}, 575\penalty0 (7784):\penalty0 622--627, November 2019.
\newblock ISSN 1476-4687.
\newblock \doi{10.1038/s41586-019-1709-y}.
\newblock URL \url{https://www.nature.com/articles/s41586-019-1709-y}.
\newblock Number: 7784 Publisher: Nature Publishing Group.

\bibitem[Quan and Loncar(2011)]{quan_deterministic_2011}
Qimin Quan and Marko Loncar.
\newblock Deterministic design of wavelength scale, ultra-high {Q} photonic
  crystal nanobeam cavities.
\newblock \emph{Optics Express}, 19\penalty0 (19):\penalty0 18529--18542,
  September 2011.
\newblock ISSN 1094-4087.
\newblock \doi{10.1364/OE.19.018529}.
\newblock URL
  \url{https://www.osapublishing.org/oe/abstract.cfm?uri=oe-19-19-18529}.
\newblock Publisher: Optical Society of America.

\bibitem[Chen et~al.(2017)Chen, Zhao, Tan, Xu, Wu, Liu, Fu, Fu, Geng, Liu, Liu,
  Tang, Li, Zhou, Sum, and Loh]{chen_chemical_2017}
Jianyi Chen, Xiaoxu Zhao, Sherman J.~R. Tan, Hai Xu, Bo~Wu, Bo~Liu, Deyi Fu,
  Wei Fu, Dechao Geng, Yanpeng Liu, Wei Liu, Wei Tang, Linjun Li, Wu~Zhou,
  Tze~Chien Sum, and Kian~Ping Loh.
\newblock Chemical {Vapor} {Deposition} of {Large}-{Size} {Monolayer} {MoSe2}
  {Crystals} on {Molten} {Glass}.
\newblock \emph{Journal of the American Chemical Society}, 139\penalty0
  (3):\penalty0 1073--1076, January 2017.
\newblock ISSN 0002-7863.
\newblock \doi{10.1021/jacs.6b12156}.
\newblock URL \url{https://doi.org/10.1021/jacs.6b12156}.
\newblock Publisher: American Chemical Society.

\bibitem[Liu et~al.(2020)Liu, Wu, Bai, Chae, Li, Wang, Hone, and
  Zhu]{liu_disassembling_2020}
Fang Liu, Wenjing Wu, Yusong Bai, Sang~Hoon Chae, Qiuyang Li, Jue Wang, James
  Hone, and X.-Y. Zhu.
\newblock Disassembling {2D} van der {Waals} crystals into macroscopic
  monolayers and reassembling into artificial lattices.
\newblock \emph{Science}, 367\penalty0 (6480):\penalty0 903--906, February
  2020.
\newblock ISSN 0036-8075, 1095-9203.
\newblock \doi{10.1126/science.aba1416}.
\newblock URL \url{https://science.sciencemag.org/content/367/6480/903}.
\newblock Publisher: American Association for the Advancement of Science
  Section: Report.

\bibitem[Yang et~al.(2015)Yang, Giessen, and Lalanne]{yang_simple_2015}
Jianji Yang, Harald Giessen, and Philippe Lalanne.
\newblock Simple {Analytical} {Expression} for the {Peak}-{Frequency} {Shifts}
  of {Plasmonic} {Resonances} for {Sensing}.
\newblock \emph{Nano Letters}, 15\penalty0 (5):\penalty0 3439--3444, May 2015.
\newblock ISSN 1530-6984.
\newblock \doi{10.1021/acs.nanolett.5b00771}.
\newblock URL \url{https://doi.org/10.1021/acs.nanolett.5b00771}.
\newblock Publisher: American Chemical Society.

\bibitem[Shahnazaryan et~al.(2017)Shahnazaryan, Iorsh, Shelykh, and
  Kyriienko]{shahnazaryan_exciton-exciton_2017}
V.~Shahnazaryan, I.~Iorsh, I.~A. Shelykh, and O.~Kyriienko.
\newblock Exciton-exciton interaction in transition-metal dichalcogenide
  monolayers.
\newblock \emph{Physical Review B}, 96\penalty0 (11):\penalty0 115409,
  September 2017.
\newblock \doi{10.1103/PhysRevB.96.115409}.
\newblock URL \url{https://link.aps.org/doi/10.1103/PhysRevB.96.115409}.
\newblock Publisher: American Physical Society.

\end{thebibliography}
\bibliographystyle{unsrtnat}
	
\end{document}